\documentclass[twocolumn,aps,prb,amsmath,amssymb,superscriptaddress]{revtex4-1}    

\usepackage{graphicx}
\usepackage{epstopdf}
\usepackage{color}
\usepackage{float}

\begin{document}

\preprint{Bi$_2$Te$_2$Se, Shekhar et al.}

\title{Evidence of surface transport and weak anti-localization in single crystal of Bi$_2$Te$_2$Se  topological insulator}

\author{Chandra Shekhar}
\email{shekhar@cpfs.mpg.de}
\author{C. E. Viol Barbosa}
\author{Binghai Yan}
\author{Siham Ouardi}
\author{W. Schnelle}
\author{Gerhard~H. Fecher}
\author{Claudia Felser}
\email{felser@cpfs.mpg.de}
\affiliation{Max Planck Institute for Chemical Physics of Solids, 01187 Dresden, Germany.}

\date{\today}

\begin{abstract}
Topological insulators are known to their metallic surface states, a result of strong-spin-orbital coupling, that show unique surface transport phenomenon. But these surface transports are buried in presence of metallic bulk conduction. We synthesized very high quality Bi$_2$Te$_2$Se single crystals by modified Bridgman method, that possess high bulk resistivity of $>$20~$\Omega$cm below 20~K, whereas the bulk is mostly inactive and surface transport dominates. Temperature dependence resistivity follows the activation law like a gap semiconductor in temperature range 20-300~K. We designed a special measurement geometry, which aims to extract the surface transport from the bulk. This special geometry is applied to measure the resistance and found that Bi$_2$Te$_2$Se single crystal exhibits a cross over from bulk to surface conduction at 20~K. Simultaneously, the material also shows strong evidence of weak anti-localization in magneto-transport due to the protection against scattering by conducting surface states. This novel simple geometry is an easy route to find the evidence of surface transport in topological insulators, which are the promising materials for future spintronic applications.

\end{abstract}



\maketitle


Topological insulators (TIs) have recently attracted significant attention since they are materialized into a new state of quantum matters, and they possess topologically protected metallic states on their edges or surfaces.~\cite{MZH10,XLQ11} These conducting SSs originate from the inversion of bulk bands due to the results of strong-spin-orbital coupling. Therefore, these novel metallic states are a subject of intensive investigations because of not only its fundamental novelty but also its high potential applications for spintronics devices ~\cite{MK07} and quantum computations.~\cite{CLK06} 
	
	In 3D TIs, the topological surface states have been successfully investigated by surface-sensitive techniques as angle-resolved photoemission spectroscopy~\cite{MN12,LB12} and scanning tunneling microscopy ~\cite{ZA10,TZ09}. However, the evidences of transport through the surface sates remain a challenge due to the presence of the parallel bulk conducting channels that usually dominate the transport properties.~\cite{NPB10,AAT11,ZR11} Various strategies to improve surface conductivity, e.g., studying thin films or nano-flacks~\cite{YSF13,BX13,BAA13,JL12,LB12,ZL12}, gating~\cite{JL12,JC11,JC10}, and doping~\cite{ZR12,AAT11,ZR11} have been employed. Among the all known TIs, the ternary \textit{tetradymite} Bi$_2$Te$_2$Se (BTS) has largely been investigated material that gives excellent performance of surface in temperature dependence resistivity~\cite{YSF13,BX13,BAA13,JL12,ZL12,LB12,JX12}. It shows large bulk resistivity due to the nearly perfect crystalline structure, and it has been predicted nearly perfect Dirac cones in terms of less entanglement of bulk and surface states ~\cite{ZL12,LB12,ZR12,JX12}. The quantum oscillations have also been observed in BTS~\cite{JX12,SJ11,ZR10} and doped BTS~\cite{ZR12,AAT11} bulk samples, and they show surfaced-dominated transport contributes up to 70$\%$ of the total conductance~\cite{JX12,AAT11}. Besides showing this metallic behavior, excellent performance of the surface states of TIs appears in magneto-conductance in effect of weak anti--localization (WAL), which is observed either thin films or nano flacks only. ~\cite{LB12,ZL12} In view of these promising advances of the BTS compound, the exploration of the topologically protected surface transport in bulk TIs appears to become true. Very recently, SmB$_6$ has been identified as a topological insulator ~\cite{ZHZ13,NX13}, that shows surface dominated transport below 4~K in resistivity, which has been measured in the specialized geometry of single crystals. ~\cite{SW13,DJK13}
		
The main goal of the present work is to explore the performance of BTS compound as TI in state of art high quality crystals. The resistance of our BST crystal shows a steep growth with decreasing temperatures and starts to saturate at T$\leq$30~K. This steep growth appears in the form of activation energy with $\Delta$ = 28.1~meV between temperature range 300-30~K, and after 30~K, it shows metallic character due to presence of topological SSs, which is the main attraction of a TI. We explored these surface states by measuring resistance in a especially designed geometry and results show a cross over from bulk to surface conduction below 20~K. Besides this, the magneto-conductance provides a clear evidence of WAL, which is also pointing towards a strong contribution of topological SSs. 


Single crystals of BTS were grown by modified Bridgeman method from high-purity (99.999$\%$) bismuth (Bi), tellurium (Te) and selenium (Se) chips/ingots. First, the elements were loaded stoichiometry ratio into a dry alumina tube and then, put into dry quartz tube. The tube was evacuated (10$^{-6}$ torr) and sealed. The sealed ampoule was loaded into the vertical Bridgeman furnace and heated slowly to form a single phase. The ampoule was heated to 800$^o$C at 60$^o$C/h and followed by 12 hours soaking. For single crystal growth, the temperature was slowly reduced by 2$^o$C/h from 800$^oC$ to 500$^o$C and annealed it for a week at 500$^o$C before cooling to room temperature. This procedure resulted in silver-colored single crystals size of 10 mm. Slow cooling rate and extra annealing procedures were followed to minimise the defects in form of the Se vacancies, which are invariably found in the Se containing \textit{tetradymite} compounds. The crystals were confirmed to be single phase and identified as having the rhombohedral Bi$_2$Te$_2$Se crystals structure by x-ray diffraction. X-ray diffraction  patterns(not shown here) of the cleaved crystals oriented with basal plane normal bisecting the incident and diffracted beam directions showed only the (006), (009), and (0012) peaks etc. (hexagonal setting), indicating that the cleaved surface is oriented perpendicular to the hexagonal c axis. X-ray revealed structure of hexagonal with space group R-3 m (No. 166). The lattice parameters were determined to be $a = 4.284 \AA$, $c = 29.669\AA$. The observed values are good agreement with those reported previously ~\cite{JC11, TLA74}.

\begin{figure}[htb]
\includegraphics[width=8.5cm]{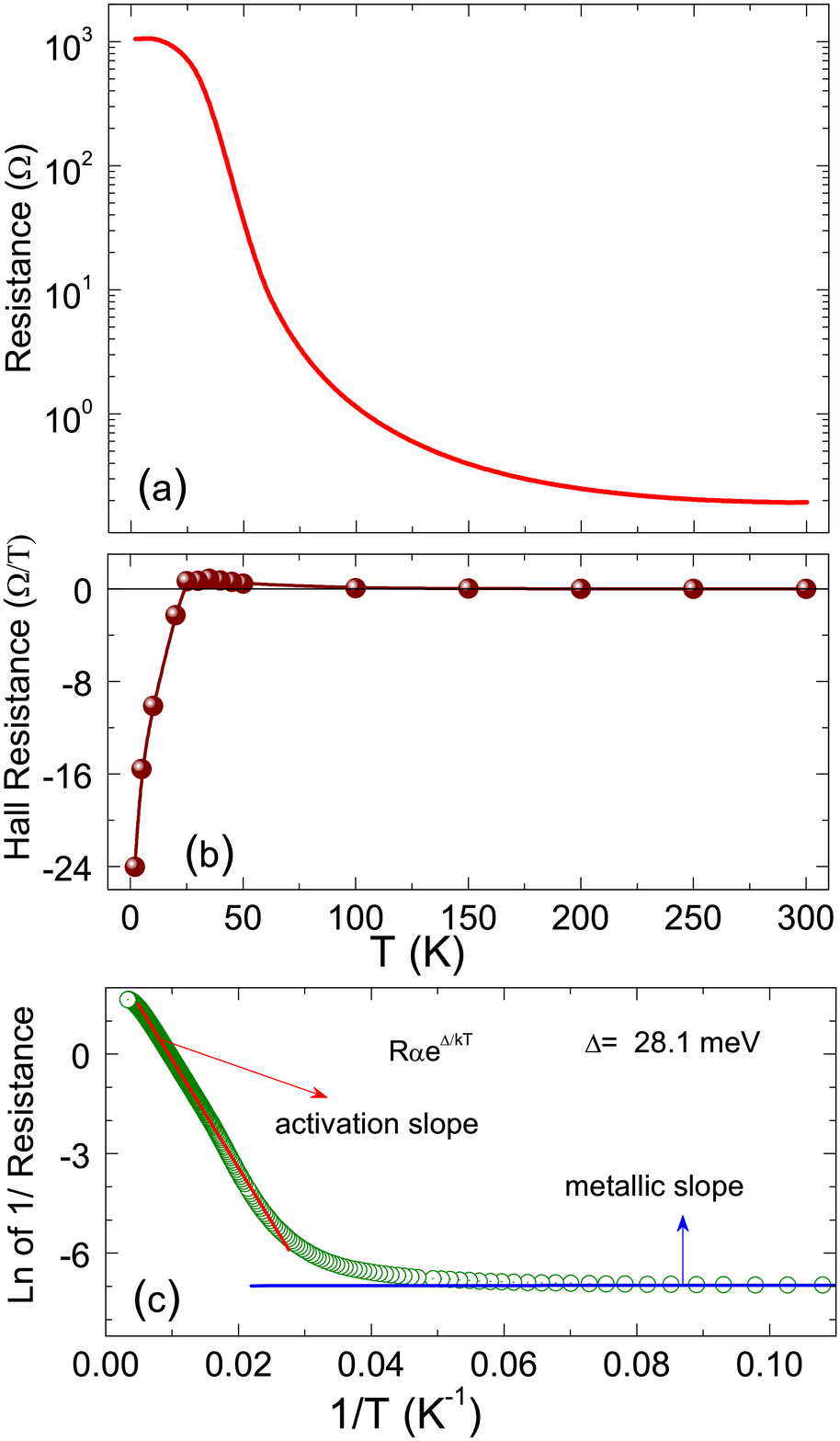}
\caption{ (Color online) Temperature dependence of (a) four-probes resistance of bulk single crystal, (b) four-probes Hall coefficients and (c) resistance follows the Arrhenius law, $R=R_0$e$^{\Delta/kT}$ in the temperature range between 300 and 30~K as indicated by the straight red line and after that it follows metallic behavior (blue line).}
\label{fig:fig1_RT}
\end{figure}

To explore the transport properties, we performed usual four-probes to measure normal resistance and five-probes Hall resistance of bulk BTS single crystal of dimensions 2.6 $\times 0.4$$\times5.2$ ($w\times d\times l$)~mm$^3$ in the temperature rages 2--300~K and field up to 9~T. Ohmic contacts were made by using conductive silver epoxy. Fig.~\ref{fig:fig1_RT} shows the temperature dependency of the electrical resistance, $R(T)$ and Hall coefficient, $R_H$(T). Over all zero field  temperature dependency of the resistance is shown in Fig.~\ref{fig:fig1_RT}(a). The magnitudes of $R$ (T) are 1~k$\Omega$ at 2~K and 0.2~k$\Omega$ at 290~K showing that a big change of more than three order of magnitudes in resistance is found and it steeply increases from temperature 300~K to ~30~K reflecting that sample possesses band gap. This saturation temperature is much similar to the previously observed in the BTS compound.~\cite{BX13,JL12,JX12,LB12,ZL12,ZR10,SJ11} After saturation, the resistance reaches its maximum point at 20~K then start to  decreases. A fitted blue line in the Fig.~\ref{fig:fig1_RT}(c) shows the metallic behavior below 20~K due to presence of metallic SSs of TIs, which are the main attraction of the TIs. In order to get broad insight of change transport mechanism, we carefully analyzed the variation of resistance in the high-temperature range by plotting $ln(1/R)$ versus ($1/T$) as shown in Fig.~\ref{fig:fig1_RT}(c). The ln$(1/R)$ from 300 to 40~K is very well fitted by activated transport model with $\Delta=28.1$~meV. This transport behavior and value of $\Delta$ are similar to the other reported BST crystals. ~\cite{ZR10,ZR12} This indicates that the large bulk resistivity of this compound can be achieved by taking stoichiometric ratio of initial elements, which avoids the manipulation about initial elemental ratio.~\cite{ZR10,ZR12,AAT11,ZR11} Fig.~\ref{fig:fig1_RT}(b) summarizes the result obtained from Hall measurements. The Hall coefficient changes sign positive to negative as temperature is changed high to low. This result is similar to other reports~\cite{ZR10, SJ11,LB12}, where negative charge carriers dominate at low temperatures. A large change in resistance relates its intrinsic behavior, however high value of Hall coefficient i.e.low carrier density relates the presence of minimum defects in sample. Most of the reduction in $R(T)$ with increasing temperature is mainly due to an increase of the carrier concentration except below 20~K, whereas charge carrier saturates. 

\begin{figure}[htb]
\includegraphics[width=8.5cm]{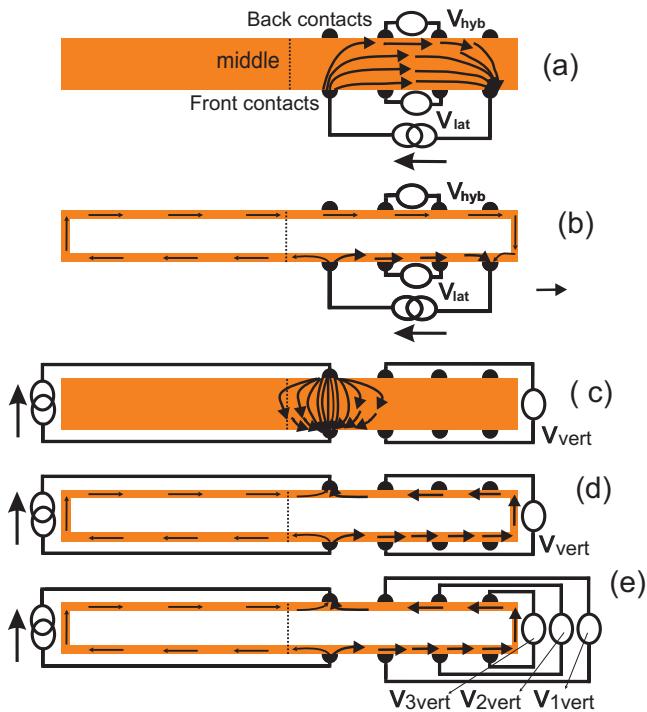}
\caption{ (Color online) A horizontal cross section of the sample along with eight coplanar electrical contacts, four on each side. (a) The current goes laterally through the bulk, and the front-side and back-side voltages give same values. (b) The bulk in (a) becomes insulating, the current force to flow along the edge (i.e. along the surface in sample) and front voltage contacts isolate from back contacts. (c) Current goes horizontally through the bulk and voltage contacts are far away from the major current path.(d) The bulk in (c) becomes insulating, the current force to flow along the edge. (e) Measure the voltage in (d) at three consecutive contacts. All these diagrams have been made by assuming the isotropic transport, where the thick and thin arrows indicate the major and minor current directions, respectively.}
\label{fig:fig2}
\end{figure}
\begin{figure}[htb]
\includegraphics[width=8.5cm]{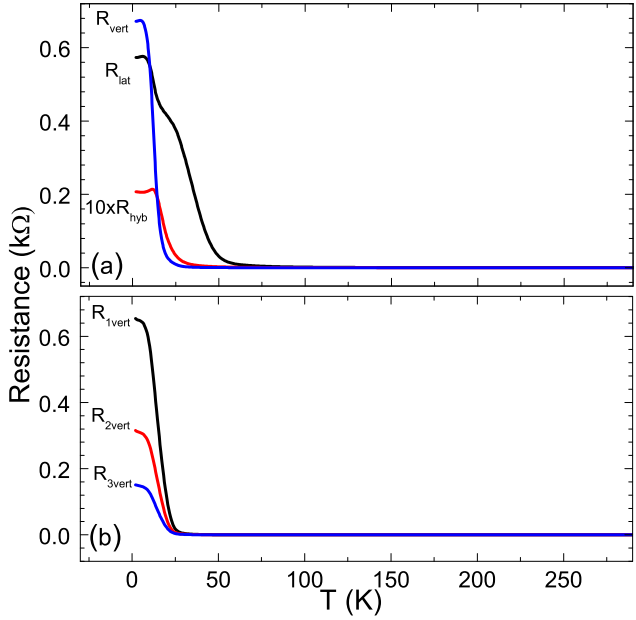}

\caption{(Color online) Temperature dependence of resistances which are measured in a special configuration (a) R$_{lat}$, R$_{hyb}$ and R$_{vert}$, (b) R$_{vert}$ at three consecutive points in the same configuration.}
\label{fig:fig3_RT}
\end{figure}

Main attraction of TIs is transport through SSs but this is usually buried in presence of bulk conduction. From Fig.~\ref{fig:fig1_RT}(a), the plotted resistance shows similar temperature dependence behavior as previously reported, which has been measured by conventional four-probes method. At temperature higher than 30K, an uniformly current passes through whole material due to bulk conducting but these currents force to flow through only surface when the bulk becomes insulating at temperature below 30K, where compound shows metallic behavior. To extract the surface transport from the bulk of well characterized crystal, we measured the four-probes resistance in a special geometry putting the voltage and current contacts at unusual position.~\cite{SW13} Here, V$_{lat}$ is normal voltage which has been measured in such a way that voltage and current contacts are lie on the same surface (front surface), V$_{hyb}$ is a bulk voltage which has been measured by putting the current and voltage contacts in opposite surface (current contacts on the front surface and voltage contacts on the back surface). However, V$_{vert}$ is a voltage which has been measured by putting one voltage and one current contacts on same surface (front surface) but other two contacts on the opposite surface (back surface). This V$_{vert}$ value is measurable only when surface is conducting otherwise this is unmeasurable due to conducting bulk. To avoid the flow of current in other directions, we made the contacts after half of the sample, whereas all the voltage contacts lie in their own respective path of majority current. Such typical contacts configurations are illustrated in Fig.~\ref{fig:fig2} and measured resistances according to these configurations are shown in the Fig.~\ref{fig:fig3_RT}. If the material is a bulk conducting, the mentioned currents path in the Figs.~\ref{fig:fig2} (a) (c) will be a major path otherwise current will follow the path as mentioned in Figs.~\ref{fig:fig2} (b) (d), which are the case of surface transport. Therefore, the measured resistances R$_lat$, R$_{hyb}$ and R$_{vert}$ are extremely different to each other and its values totally depend upon their current path. In the case of bulk conduction, R$_{lat}$ and R$_{hyb}$ should be equal due to same amount of current passes between V$_{lat}$ and V$_{hyb}$ probes but R$_{vert}$ is very low because of only negligible amount of currents reach between V$_{vert}$ probes. The magnitude of R$_{lat}$ $>$ R$_{hyb}$ $>$ R$_{vert}$  are found, which are expected as a bulk conduction at high temperature. But in case of surface conduction or TI case below 30~K , where current force to flow through the surface, get R$_{lat}$ lager than R$_{hyb}$ because V$_{lat}$ and V$_{hyb}$ probes lie in major current path on the front surface and minor current path on the back surface respectively. However, R$_{vert}$ is highest because V$_{vert}$ contacts lie on the majority current path, which is also longer than the path of the V$_lat$ probes.~\cite{SW13} The measured data are shown in Fig.~\ref{fig:fig3_RT}, which are justified according to surface and bulk transport. To explore resistance of R$_{vert}$ further, we measured it at three consecutive points as shown in the Fig.~\ref{fig:fig2}(e). The measured resistances R$_{1vert}$, R$_{2vert}$ and R$_{3vert}$ are displayed in the Fig.~\ref{fig:fig3_RT}(b). In these values, R$_{1vert}$ is highest and R$_{3vert}$ is lowest, which are again with respect to their current path that means R$_{1vert}$ covers largest path but R$_{3vert}$ covers least path according to surface transport.

\begin{figure}[htb]
\includegraphics[width=8.5cm]{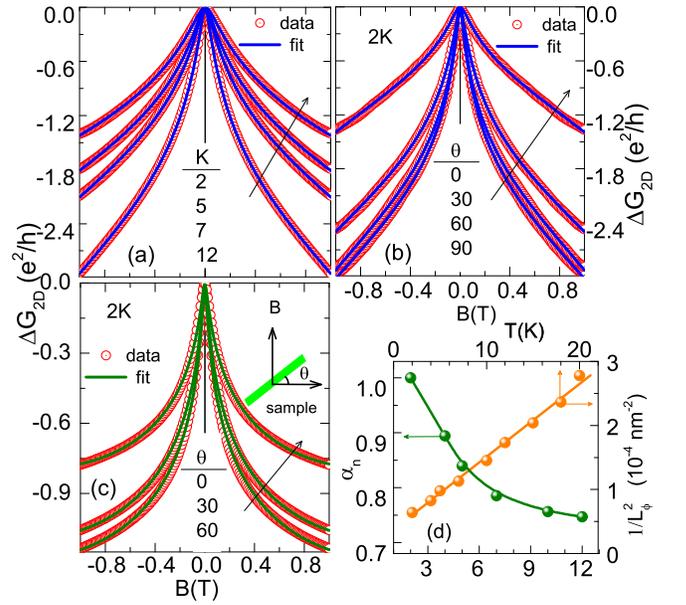}
\caption{(Color online) (a) Temperature dependence of conductivity. (b) Conductivity measured in titled B fields, where, blue line is fitting by HLN 2D-model of equation (2). (c) Perfect 2D data in tilted B fields are obtained by subtracting tilted data with $\theta=90$, where, green line is fitting by only first term of HLN 2D-model of equation (2) (d) Temperature dependence of normalized $\alpha_n=\alpha/\alpha_{2K}$ (left axis) and $1/L^2_\Phi$ (right axis).}
\label{fig:Conductance-2K-12K}
\end{figure}

In order to get better insight into transport properties in the low temperature range, we investigated the magneto-conductance in tilted field. For this purpose, we choose temperature range in the metallic region, which has already been explained in the Fig. ~\ref{fig:fig1_RT} and Fig. ~\ref{fig:fig3_RT} . The nature and values of the magneto-conductance strongly depend on the applied fields as shown in Fig.~\ref{fig:Conductance-2K-12K}. $\Delta G_{2D}$ represents the change of the 2D conductance ($G_{2D}$) with fields and is estimated as $G_{2D}=G(d/w)$, where $G=1/R$ and $d$ is distance between voltage probes. The symbols in Fig.~\ref{fig:Conductance-2K-12K}(a) represent the experimental data. The sharp increase of the conductivity at low fields, forming a steep cusp, is a well-known signature of the weak anti-localization (WAL). The WAL effect arises in a phase coherent conductor when a destructive interference is formed between two time-reversed electron paths. This destructive interference inhibits elastic backscattering, whereby increases the conductivity . 


This WAL effect is found in only thin films or nano-flacks of TIs~\cite{ZL12,LB12,ML12}, and originates from strong spin--orbit--coupling in the bulk resulting spin--momentum locking in the topological SSs.~\cite{KN07} The response of the conductance to the magnetic field in a 2D-system with WAL effect can  be quantified by using a well known Hikami--Larkin--Nagaoka (HLN) model. ~\cite{SH80} 

\[\Delta G_{2D}^{HLN}=G_{2D}(B)-G_{2D}(0)=\]
\[\alpha\frac{e^2}{2\pi h}\left[\Psi\left(\frac{1}{2}+\frac{B_\phi}{B})\right)-ln\left(\frac{B_\phi}{B}\right)\right]\]
\[+\alpha\frac{e^2}{\pi h}\left[\Psi\left(\frac{1}{2}+\frac{B_{so}+B_{e}}{B})\right)-ln\left(\frac{B_{so}+B_{e}}{B}\right)\right]\]
\[+\alpha\frac{3e^2}{2\pi h}\left[\Psi\left(\frac{1}{2}+\frac{(4/3)B_{so}+B_{e}}{B}\right)-ln\left(\frac{(4/3)B_{so}+B_{e}}{B}\right)\right]~~~~~~~~~~~~(1)\]

where, $B_i$ are the characteristic fields of each respective scattering channel ($i=\phi,so,e$) given by $Bi=\hbar/(4eL^2_i)$. $L_\phi$ is the phase coherence length (traveled length without coherency break), $L_{so}$ is the spin-orbit scattering length, and $L_e$ is the elastic scattering length (or the mean free path). At low temperatures and small fields, coherence length is the longest among these three lengths, therefore, the phase coherence scattering dominates all over existing scattering.~\cite{BAA13} $\Psi$ is digamma function.  At high fields, the $L_{so}$ and $L_e$ lengths become prominent and yield characteristic fields of the order of several tesla. It can easily be approximated latter two terms containing spin-orbit scattering and elastic scattering into a $B^2$ term. Therefore, this approximation leads to

\[\Delta G_{2D}=\]
\[\alpha\frac{e^2}{2\pi h}\left[\Psi\left(\frac{1}{2}+\frac{B_\phi}{B})\right)-ln\left(\frac{B_\phi}{B}\right)\right]-cB^2~~~~~~~~~(2)\]

This has already been applied earlier.~\cite{BAA13} Here, $\alpha$ indicates that the types of localization and is equal to -1 for WAL effect in a perfect 2D system. But in our case, the  fitted values of $\alpha$ are 3.1--2.3 in temperature range 2--20~K. We considered that highest conduction through the surface states occurs at 2~K, and took the value of $\alpha$ at 2~K as a reference. The normalized values, $\alpha_n$ ($\alpha$/$\alpha_{2K}$), range  from 1.0--0.75 between 2-12~K as shown in Fig.~\ref{fig:Conductance-2K-12K} (d). These values decrease with increasing temperature and reveals that the surface contributions decrease with increasing temperature. Other fitting parameter is $L_\phi$, which is directly related to $B_\phi$. The 2D nature of conductivity shows more qualitative from temperature dependence of $L_\phi$ that varies with $T^{-0.5}$ for inelastic electro--electron interaction.~\cite{GB84,BLA82} From the Fig.~\ref{fig:Conductance-2K-12K}(d), the temperature dependence of $1/L^2_\phi$ shows a linear behavior, indicating that the dominant inelastic  electron--electron scattering in the surface conducting channels of our BST bulk single crystal.~\cite{BX13,JL12,LB12,BAA13} The depth and magnitude of the conductivity also decrease as  increases temperature (see Fig.~\ref{fig:Conductance-2K-12K} (a)) due to the decrease of the $L_\phi$ and these maintain up to 147~K.~\cite{BAA13} 
	Main outcome of HLN fit is the values of $\alpha$ and $L_\phi$, which relate the number of active conducting channels and distance traveled by charge carriers before dephasing, respectively. For WAL effect of any TI, the value of $\alpha$ is experimentally found around -0.5 per conducting channel but there is no any theoretically predicted value of $~\alpha$.~\cite{BX13,JL12,LB12,BAA13} For more clarification about perfect 2D WAL effect in our bulk sample, the conductivity measured in titled magnetic fields at 2~K, as shown in Fig~\ref{fig:Conductance-2K-12K}(b). At $\theta=0^o$ i.e. B in sample plane	along current direction, the conductivity dip feature still appears. If the conduction is only through SSs, this dip feature of conductivity vanishes at $\theta=0^o$ but this is not the case hare.~\cite{HTH11} Besides 2D WAL in SSs of TIs, other WAL also exit and this might be due to bulk origin of 3D WAL.~\cite{JL12} To obtain only 2D conductivity, the 3D conductivity at $\theta=0^o$ was subtracted from the measured conductivity at other angle. The resultant data are plotted, and fitted by HLN theory with only first term in Fig.~\ref{fig:Conductance-2K-12K}(c). This fit yields $\alpha= 1.02$ and $L_\phi$= 212~nm at $\theta=0^o$, which are perfectly matched with other reports and indicate the 2D transport through SSs.~\cite{HTH11,JL12}

The observed values of $\alpha$ in literatures cover a wide range from -0.4 to -2.5, suggesting that it depends on (i) sample thickness ~\cite{BAA13,YSK11}, (ii) presence of bulk conductivity gating~\cite{JL12,JC11,JC10}, and (iii) substrate.~\cite{HTH11} The thin films of TIs, which have been prepared on substrate with different route, show the value of $\alpha$ around~0.5 ~\cite{HTH11,ML11,JC10}. But it also depends on the number of conduction channels, and $\alpha$ can even become -1 when both the top and the bottom surface of a crystals or films contribute independently ~\cite{JGC11,BX13}. However, values of $\alpha$ have also been tuned between 0.5 to 1.0 by applying gate voltage.~\cite{JC11a,JC10} Kim et al. have been observed WAL effect in a wide range of thicknesses from 3.0~nm to 0.17~mm on Bi$_2$Se$_3$ films, where values of $\alpha$ show increasing trend with thickness.~\cite{YSK11} This higher value can be explained by assuming the conduction through multiple channels i.e. some bulk channels conduct together with SSs of TIs.~\cite{HZL11} This is a plausible reason for higher values of $\alpha$ in our BTS single crystal. Another important point should be noted here about the values of $L_\phi$ from the literatures, which are usually larger than the film thickness and this makes the whole sample effectively more 2D.~\cite{HTH11,ML11} i.e. sample itself, including the bulk and SSs, acts as a 2D transport system. This type of 2D transport behavior cannot be completely contributed from the surface states. Our fitted $L_\phi$ varies from 125 to 60~nm in the temperature range 2-20~K, which are much smaller than the thickness (0.4~mm) of the sample and indicates that more 2D. Therefore, unusually measured resistance and strong WAL effects are a strong evidence of surface transport in BTS crystal.



In summary, we found the evidence of surface transport in highly bulk resistive single crystal of Bi$_2$Te$_2$Se topological insulator. Temperature below 30~K, the resistivity shows metallic behavior due to the presence of conducting surface states. Resistance measured by special geometry differentiates the transport between bulk and surface. The magneto-conductance gives the evidence of week anti-localization effect, which is mainly contributed by 2D surface states. The measured resistances  with and without field together indicate that the surface transport dominate over bulk transport below 30~K.  Therefore, this material opens new opportunities for further physical investigations and wide range applications of topological insulators. But, the surface transports still have a challenge at room temperatures where bulk transport dominates. 

\bigskip

The financial support by the DfG (P 1872.3-A in FOR 1464 ASPIMATT) is gratefully acknowledged.
BY acknowledge the support of Max-Planck-Institut für Physik komplexer Systeme in Dresden, Germany.


\end{document}